\documentclass[12pt]{iopart}					
\usepackage{graphicx}						
\usepackage{amssymb}

\begin{document}

\title{Anomalous Effects from Dipole~-~Environment Quantum Entanglement}

\author{Elio B. Porcelli$^1$ and Victo S. Filho$^1$} 
\address{$^1$H4D Scientific Research Laboratory, S\~ao Paulo, S\~ao Paulo State, 04674-225, Brazil}
\ead{elioporcelli@h4dscientific.com}
\begin{abstract}
In this work, we analyze anomalous effects observed in the operation of two different 
technological devices: a magnetic core and a parallel plate (symmetrical or asymmetrical) capacitor.
From experimental measurements on both devices, we detected small raised anomalous forces that 
cannot be explained by known interactions in the traditional theories. As the variations of device 
inertia have not been completely understood by means of current theories, we here propose a theoretical 
framework in which the anomalous effects can consistently be explained by a preexisting state 
of quantum entanglement between the external environment and either magnetic dipoles of magnetic cores 
or electric dipoles of capacitors, so that the effects would be manifested by the application of a 
strong magnetic field on the former or an intense electric field on the latter.  
The values of the macroscopic observables calculated in such a theoretical framework revealed good 
agreement with the experimental measurements performed in both cases, so that the non-locality hypothesis 
based on the generalized quantum correlation between dipoles and environment is consistent as explanation 
for the anomalous effects observed. The control and enhancement of the effect can allow the future 
viability of a new technology based on electric propulsion of rockets and aircrafts. 
\end{abstract}

\maketitle

\section{Introduction}

\begin{figure}[ht]
\begin{center}
\includegraphics[width=13cm]{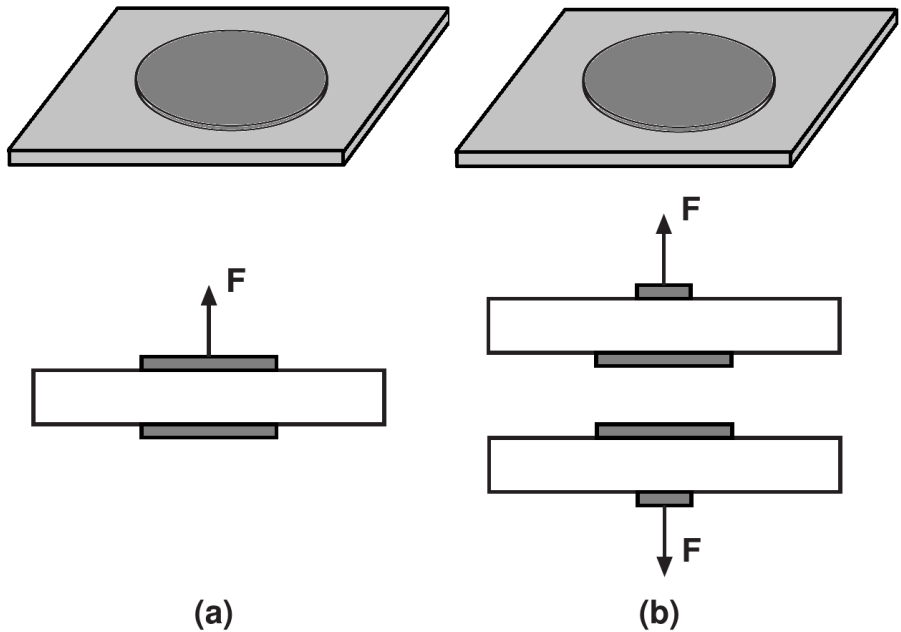}
\caption[dummy0]{Scheme of capacitor devices, constituted by two rounded parallel metallic plates and dielectric 
material between them. In the symmetrical case, it raises an upward force in the plates and in the asymmetrical capacitors 
it appears from the  small to the large plate, as indicated at the right side of the picture. At the left frame (a), 
we see an example of symmetrical capacitor and at the right frame (b) the corresponding asymmetrical capacitor in 
any possible vertical position (up or down) of the electrodes (as seen in the perfil draw). In those cases of asymmetrical 
capacitors, the anomalous force appears from the large to the small plate, as indicated in the perfil draw. In both cases, 
the plates are parallel to the ground and when the power source is turned on it is observed the anomalous force on them.} 
\label{fig1}
\end{center}
\end{figure}

We have studied a lot of very interesting and intriguing experimental results reported in several 
works~\cite{Brown1,Brown2,Williams,Manitoba,Mahood,Musha0,Musha,Fazi,Woodward1,Woodward2,Woodward3,Modanese} 
concerning to anomalous effects mainly associated with the operation of traditional high-energy capacitor 
devices. The anomaly occurs in inertia measurements and it presents a more intense magnitude of the raised force 
in experiments involving asymmetric capacitor devices~\cite{Martins,Canning1,Canning2,NASA} when they are subjected 
to high voltage. Such an anomaly was named as Biefeld-Brown effect (BB effect) due to the pioneer works from 1920 by 
Thomas Townsend Brown and Paul Alfred Biefeld in their experiments involving capacitors, in which they observed 
an upward force or thrust acting from the larger electrode toward the small one of the device~\cite{Brown1,Brown2}. 
In Fig.~\ref{fig1}, we schematically show the phenomenon observed for symmetrical and asymmetrical capacitors.  

The phenomenon was first verified in a patent published in 1928~\cite{Brown1} and it has been intensively studied since 
then, including other many experimental device configurations, as superconductor apparatus~\cite{Poher}, magnetic cores 
or flying devices~\cite{FungChung}. The anomalous forces that appear are very weak, as verified in the works cited above 
and also in our earlier works~\cite{Elio1,Elio2,Elio3}. The weakness of the anomalous force was observed even in the 
experiments with very higher voltages up to 100 kV, reported in Ref.~\cite{Manitoba}. In those experiments, it was 
observed a reduction of their apparent weight even changing the polarity and, despite of its weak magnitude, the 
observed force consistently appeared and acted on the charged capacitor in a vertical direction when their plates were 
parallel to the ground. Even so, although a lot of measurements of inertia in capacitors operating in high voltage 
have supported the existence of their apparent weight reduction~\cite{Mahood,Woodward1,Woodward2,Woodward3}, there has 
still been a reasonable controversy, mainly due to the weakness of the force generated on the devices. However, the studies 
have been more and more conclusive in the last years about the existence of the phenomenon due to an increasingly relevant 
research work by means of a lot of recent publications~\cite{Musha,Modanese,Martins,Ianconescu}. Moreover our early works 
based on huge automatic collected data by using very advanced accelerometers~\cite{Elio1,Elio2,Elio3} also corroborate the 
existence of Biefeld-Brown effect. 

The importance of verification of its existence is doubtless unquestionable because it will probably provide us with a new 
valuable resource for technological applications as the propulsion ones. The concept of propulsion without propellant has 
as one version the possibility of manipulating empty space itself (the quantum vacuum) so that one can provide 
a thrust on space rockets~\cite{Puthoff98}. In Refs.~\cite{Puthoff98,Puthoff02}, it is analyzed this hypothesis of field 
propulsion, emphasizing the possibility that interactions between matter and vacuum might be engineered for space 
applications, so that it would be neither so unfeasible from a fundamental point of view nor unviable from 
a technical point of view, although it is certainly restricted by strong constraints.  

Although more and more the phenomenon has been experimentally verified, the theoretical explanation for BB effect remains 
open and it has been pursued by many researchers. Possible theoretical explanations proposed in the literature for the 
phenomenon include the Law of Coulomb, electric wind~\cite{Li,Zhao}, Corona effect~\cite{Zhang} or fluctuations of the 
vacuum~\cite{Puthoff89,Puthoff91,Puthoff90,Haisch}. However, as the internal net resultant force is zero in capacitors, 
the Law of Coulomb could not theoretically explain the effect. Besides, a possible force between the 
Earth and the capacitors would not have the right upward direction and its magnitude was also several orders of 
magnitude higher than that one predicted by the traditional electromagnetic theory. In fact, the weak raised 
force was dependent on the magnitude of the electrical potential energy stored on the capacitor. Corona effect 
or electric wind~\cite{Antano,Moller,Tajmar} could not explain the magnitude of the anomalous force as well. 
Experiments in which electric or ionic wind was considered as possible theoretical explanation were 
realized in Refs.~\cite{Moller,Tajmar}. Besides, some experiments with capacitors were made with high 
insulation or in vacuum, as in Honda laboratory~\cite{Musha0,Musha}, not indicating the correct magnitude 
of the raised forces predicted by such a hypothesis. Rather measurements performed in the vacuum showed 
that the effect persisted even so~\cite{Brown2,Antano} and the experiments described in 
Refs.~\cite{Manitoba,Musha0,Musha} showed that the insulating materials around the 
capacitor devices could neither reduce nor eliminate the effect.   

Beside Law of Coulomb and wind effect, the remaining alternative hypothesis that has been proposed 
in order to explain BB effect is the zero-point vacuum field theory (ZPFT). Beside the works earlier cited, 
such a hypothesis was successfully considered in Ref.~\cite{Musha0}, that is, it had relative 
success in explaining BB effect by means of the interaction between the vacuum fluctuations and the high 
potential electric field provided by the capacitor. For a capacitor sample of mass 62 g reported in 
Ref.~\cite{Musha0}, the theory revealed a variation $\Delta M$ = 0.31 g for DC 18 kV, which was close 
to the experimental value reported in the work, that is, $\Delta M$ = (~0.29 $\pm$ 0.17~)~g. However, 
one realizes that the average value of weight loss calculated from ZPFT presented a relatively significant 
percentual error (close to $6.9\%$), so that there is still margin to a better theoretical explanation 
of BB effect. Besides, as the error bar was large in that work, experimental measurements must also be improved.  

In order to pursue a better theoretical explanation for the phenomenon, we proposed in Refs.~\cite{Elio1,Elio2,Elio3} 
an empirical formulation that showed to be possible describing with very good accuracy the anomalous effect by the 
macroscopic manifestation of the microscopic entanglement of all electric dipoles on the dielectric from the same 
ancient idea used by Clausius and Mossotti in their works~\cite{Clausius,Mossotti,Rysselberghe}, based on a 
straightforward relationship between microscopic and macroscopic observables. However, it was still missing 
a theoretical basis in order to justify such an empirical hypothesis.  

Based on this motivation, we here investigated BB effect now under two perspectives: in a first one, we 
investigate the theoretical explanation of the anomaly from a more fundamental point of view. Besides, 
in a second investigation, in order to reinforce the consistency of the theory, we show that the idea 
is more profound and general that the proposal one already described only for capacitors. In fact, we 
implemented a new experimental setting in which by applying strong magnetic fields measurements of forces 
induced in magnetic cores were performed. In this way, one could also verify the idea of relationship 
between microscopic properties from magnetic dipoles in the devices and their macroscopic effects on 
the macroscopic observables like the magnetic susceptibility. Thus it was also possible to analyze 
the Clausius-Mossotti hypothesis also in this new case. 

In the following, we present our theoretical model based on an alternative explanation of the 
phenomenon by considering a more general point of view. Basically, we describe the principle of the 
alternative scenario based on the manifestation of macroscopic observables by means of microscopic 
ones. Afterward, we describe the experiments performed with capacitors and magnetic cores and we 
detail our experimental setups and measurements concerning to weight variations of the devices. 
We also compared our calculations with the experimental measurements and discuss the accuracy of 
the theoretical results. Finally, in the last section we present our conclusions and final remarks. 
\section{Theoretical Description}
\label{sec:3}
 
\subsection{Magnetic Dipole Forces and Generalized Quantum Entanglement}

Our method of calculation considers, for the microscopic constituents of a physical system, the property of 
preexisting state of generalized quantum entanglement, so that its existence in extreme conditions would 
be manifested and its effects could occur and be observed. Rather we assert that the effects can be 
calculated, in order to predict the values of macroscopic observables associated to the systems. 
As known from literature~\cite{EPR,Bohm,Bell,Zeilinger,Lengeler}, the quantum entanglement is the 
consequence of an interaction between two~\cite{EPR,Bohm,Bell} or more particles in an arbitrary 
past time, so that subsequently each particle in the combined state keeps the information from each 
other, even for large distances, a feature of nonlocal phenomena~\cite{Zeilinger,Lengeler}. 

The issue concerning to effects of generalized quantum entanglements in systems of huge number of particles 
is an important point to be analyzed if we desire to understand and even quantify the influence of their 
effects on our macroscopic world. This point has been avoided in a lot of papers for a long time, as cited in 
Ref.~\cite{Penrose}, in which an analysis of quantum entanglement phenomenon is considered 
in all of its features and consequences. In this reference, Penrose claims that there are two mysteries 
associated with the entanglement phenomenon: first, the quantum entanglement phenomenon itself; second, it is 
the issue of the absence of the phenomenon in our daily world. Such an important point in the author's discussion 
- that issue of the absence of verification of the phenomenon in our real world - means that the physicists always 
assume that these supposed entanglements with the outside world can be ignored, a vision that seems to rely on the 
idea that the effects of entanglements will somehow cancel each other so that they do not need to be considered 
in practice, in any actual situation. However, the author argues that one mystery or puzzle of entanglement 
consists in its tendency to spread. It would seem that eventually every particle in the universe 
should become entangled with every other or even they would already be all entangled with each other. 
This point is reinforced in a recent work~\cite{Buniy}. In fact, entanglement is an ubiquitous phenomenon 
and the huge majority of quantum states are actually entangled, so it becomes very hard to understand why 
we barely notice its effects in our direct experience of the world. According to the author, it seems that 
from that point of view in which these supposed entanglements with the outside world could be ignored is 
based on the idea that the uncountable number of quantum entanglements among lots of particles in 
the systems seems somehow average out so that it does not need to be considered in practice, in any actual 
situation. 
 
We also agree that such a vision is right in many systems, but it is merely partial and in some cases  
it hides the general behavior or effects in a system with many bodies. So, we here hypothetically suppose the 
arbitrary idea that all microscopic constituents of physical systems are quantum entangled, as is the case already 
reported in literature~\cite{Buniy}. The generalized quantum entanglement could be observed as a macroscopic 
manifestation revealed by means of its possible connection with the gravity~\cite{Elio4}. Despite the effects 
of this property cannot be easily demonstrated or even measured in case of trivial physical systems, we assert that 
in specific conditions such global effects can be detected. 
In the case of magnetic cores, it is needed the application of strong magnetic fields in order to realize the 
presence of the generalized quantum entanglement among the dipoles of the dielectric and the external environment.
In the case of capacitors, it is needed the application of high voltage in order to realize the presence 
and or the effects of the generalized quantum entanglement among the dipoles of the dielectric and 
the external environment. The conjectures of preexisting quantum entanglement and its influence on the 
macroscopic observables of the system can be checked in a lot of physical systems, as the system 
constituted by magnetic  dipoles, analyzed in this paper as a possible case of application.
 
Despite the real complexity of the calculations in systems of many entangled particles, some macroscopic 
observables can offer a real possibility of successfully calculating their physical values in an easier way, 
with a relatively good accuracy, as we can see in the formalism that we show from now on. 

Let us consider an entangled system for two magnetic dipoles (N=2) of magnetic dipole moments $M_1$ and $M_2$ 
separated by the vector $\vec{r}$. The magnetic dipole moment $M_i$ at site $i$ is related to the spin $\sigma_i$
by~\cite{Amico}
\begin{equation}
	M_{ik} =  \mu_i g_k \sigma_{ik},  
\end{equation}
in which $k$ is the index representing the components x, y and z, $\mu_i$ is the magnetic moment of the spin $i$, 
$g_k$ are the elements of the anisotropic g-factor matrix (with $g_{x}=g_{y}=g_\perp\ne g_z$) and $\sigma_{ik}$ 
are the classical Ising spins that can take the values $\pm 1$. In a first approximated classical calculation in 
which $g_x=g_y=0$ such a system can be described by the hamiltonian showed in Ref.~\cite{Ghosh}:
\begin{equation}
H = - \sum_{i,j}^{N} J_{ij}\mu_i \sigma_{iz} \mu_j \sigma_{jz}, 
\end{equation}
in which $J_{ij}$ is the operator spin total of the dipole field coupling to the moment of spin $i$ and acting only 
along the Ising axis. This operator assimilates some numerical factors and falls off as $\frac{1}{r^3}$. 
The moments of spin in the hamiltonian are $\mu_i$ and $\mu_j$. 

Let us assume a negligible magnetic coupling between the two dipoles. Without loss of generality, we also 
assume that both components of that model have same dipole moment, that is, 
	\[
	|\mu_i| = |\mu_j|, 
\]
but each of them is coupled with a different local external magnetic field aligned in the z (vertical) direction, 
as similarly shown in EPRB experiment~\cite{Bohm,Bell}.

The dipole moments can be oriented along ( $| \ 0 \ \rangle$ ) or against ( $| \ 1 \ \rangle$ ) the external 
fields. In this condition, the following two entangled states for the pair of dipoles can be obtained: 
\begin{equation}
| \Psi_2 \rangle = \frac{| 01 \rangle - | 10 \rangle}{\sqrt{2}}
\label{eq4}
\end{equation}
and 
\begin{equation}
| \Psi_3 \rangle = \frac{| 01 \rangle + | 10 \rangle}{\sqrt{2}}.
\label{eq5}
\end{equation}

It means that some change in the local external field (e.g. intensity) applied in a dipole of the pair can change 
the dipole moment of another one. The total dipolar moment of the system must be conserved accordingly 
and the dipole moment variation means that a real force can be exchanged between them, but this kind of interaction 
cannot be taken in place by local forces because we had assumed before a negligible magnetic coupling. The intensity 
of the nonlocal force exchanged from a dipole to another one is directly proportional to its transition energy E:
\begin{equation}
E = - \vec{\mu} \cdot \vec{B} ,
\end{equation}
in which $\vec{\mu}$ is the magnetic dipole moment and $\vec{B}$ is the local external magnetic field in the site 
$i$ where the magnetic dipole is localized.

This argument can be generalized accordingly and we can consider a myriad of magnetic dipoles coupled via quantum 
entanglement between the magnetic core where the strong magnetic field is applied and the environment. 
The big challenge to use the quantum mechanics framework for a calculation of the nonlocal force intensity 
is the complexity of the systems composed of a myriad of magnetic dipoles with possible different dipole moments, 
a considerable magnetic coupling between them and the thermal effects.

But everyday more and more academical works~\cite{Amico,Vedral1} have considered studies concerning to quantum 
entanglement for many bodies or quantum information in macroscopic scale. In particular, there are works~\cite{Vedral} 
consolidating the argument that the quantum entanglement is crucial for explaining a lot of macroscopic properties in 
high temperatures. 

Besides, as an additional support to our theory, in Refs.~\cite{Vedral,Ghosh} it was reported that in a simulation 
of 400 magnetic dipoles the entanglement among them could explain deviations from Curie's Law, in a system composed 
by $10^{20}$ magnetic dipoles in temperatures close to 0~K. So, the authors explained a macroscopic property of a 
small bulk, e. g., the behavior of the magnetic susceptibility versus temperature of the bulk for $10^{20}$ 
magnetic dipoles in the salt, by means of a simulation with maximum number of 400 spins (8x10$^4$ pair-wise 
interactions). Above an arbitrary temperature, the result obtained using the quantum hamiltonian is the same 
one obtained using a classical hamiltonian, so that above this temperature the Law of Curie is valid. 

In summary, it is worth to emphasize that all of those articles corroborate the hypothesis of manifestations of quantum 
entanglement for many particles in the macroscopic level~\cite{Amico,Vedral1,Vedral,Ghosh}. It is also important to stress  
that such a feature allows us to use the classical formulation in our calculations. 

The behavior of the magnetic susceptibility $\chi_m$ in low temperatures (near 0 K) cannot be explained if the 
wide entanglement in spins is not considered accordingly~\cite{Ghosh}. The magnetic susceptibility curve 
considering the quantum entanglement contributions follows the Curie law for high temperatures (T $>$ 0 K), 
hiding the quantum effects. 

Nowadays there are some important works~\cite{Marcin,Wu,Toth,Dowling,Wang,Hao,Chakraborty} that show the 
existence of a macroscopic observable revealing quantum entanglement between individual spins in a solid, 
a phenomenon called entanglement witness. In Ref.~\cite{Marcin}, the entanglement  witness is shown as 
being more general (in the sense that it is not only valid for special materials), associating some 
macroscopic observables  such as magnetic susceptibility $\chi_m$ with spin entanglement between individual 
constituents of a solid. It was proposed a macroscopic  quantum complementary relation basically between 
magnetization $M$, representing local properties, and magnetic susceptibility $\chi_m$,  representing 
nonlocal properties. By defining for the system of $N$ spins of an arbitrary spin length s in a lattice 
the quantities: 
\begin{equation}
Q_{nl} = 1 - \frac{k T \bar{\chi}}{Ns}  
\end{equation}
and 
\begin{equation}
Q_l =  \frac{\langle \vec{M} \rangle^2}{N^2s^2},  
\end{equation}
in which $\vec{M}$ is the magnetization vector, $k$ is the Boltzmann constant, $T$ the temperature and 
the susceptibility $\bar{\chi}$ is defined as $\bar{\chi} \equiv \chi_x + \chi_y + \chi_z$, then it was 
shown in Ref.~\cite{Marcin} that one has:
\begin{equation}
Q_{nl} + Q_l \le 1.  
\label{rel} 
\end{equation}
Such quantities have specific meanings, that is, $Q_{nl}$ represents the quantum correlations between the 
spins in the solid (nonlocal properties) and $Q_l$ represents the local properties of individual spins.  


The hypothesis of preexisting state of quantum entanglement indicates that there are no isolated systems, 
thus the magnetic core and the environment around it are both part of the same system where 
the inequality (\ref{rel}) can be considered accordingly. In other words, if one quantity increases then 
the corresponding counterpart quantity has to decrease. If $Q_l$ increases and $Q_{nl}$ decreases in 
the magnetic core, $Q_{nl}$ decreases and $Q_l$ increases in the environment and vice-versa. 
This is the same framework described before involving a simple system with two entangled magnetic dipoles. If we 
increase the intensity of a magnetic field ($Q_l$) applied in one then the nonlocal effects ($Q_{nl}$) must increase 
in the other. By nonlocal effects one means nonlocal forces. 

The conclusion after these arguments is that the calculation of the magnitude of the nonlocal force between a magnetic 
core and the environment around it can involve the magnetic susceptibility $\chi_m$ and magnetization $M$ although 
these macroscopic observables are classical quantities. This chain of thoughts yields a new equation postulated according 
to the correct dimensional analysis and based on the calculation of the magnetic force of solenoids~\cite{Craig} as follows:
\begin{equation}
	F = \frac{0.102}{16\pi^2} \frac{S B I}{\theta} , 
	\label{forcedip}
\end{equation}
in which the force $F$ is in units of Kgf, $S$ is the area of circular surface of the cylinder core, $B$ is the magnetic 
field generated by the electric current flowing in the turns of wire of the solenoid around the cylinder core, $I$ is the 
electric current flowing in the solenoid and $\theta$ is the cylinder core radius. The calculation of force $F$  
via equation (\ref{forcedip}) is valid for a solenoid with a cylinder ferromagnetic core where the magnetic field lines are 
parallel to the symmetry axis of the cylinder. It is important to report that the term 
\begin{equation}
E_m = S B I 	
\end{equation}
in Eq.~(\ref{forcedip}) amounts to the summation of energy eigenvalues of all magnetic dipoles in the core. 
A numeric term $\left(\frac{1}{16.\pi^2}\right)$ of the force equation 
(\ref{forcedip}) is adopted to well determine the magnitude of the force and it seems to be related to the constant term 
used in the Schr\"ondiger equation for the total energy calculation of the multi-particle system.

The equation for the magnetic field magnitude calculation~\cite{Craig} can be represented as follows:
\begin{equation}
	B = \frac{\mu_0 \mu_r N I }{L} ,	
\label{forcedip1}
\end{equation}
in which  $B$ is the magnetic field in units of T, $L$ is the solenoid length, $N$ is the number of turns of 
wire around the cylinder magnetic core, $I$ is the electric current, $\mu_0$ is the vacuum permeability, $\mu_r$ 
is the relative permeability of the solenoid core material and it is directly related to the magnetic 
susceptibility $\chi_m$ via the relation $\mu_r = 1+\chi_m$. 

As we will see in this work, the magnitude of the nonlocal force F calculated regarding the equation (\ref{forcedip}) 
is in relatively good quantitative accordance with the experimental results also described in this work where the 
magnetic interaction between the energized solenoid and the environment around it really became negligible 
considering the magnetic shield and other procedures adopted accordingly. It seems that the hypothesis of 
nonlocal interaction explains why the solenoid always loses weight regardless the magnetic field in terms 
of its upward or downward direction. The symmetry axis of the cylindric magnetic core of the solenoid 
always were in the vertical position in our experiments.


When a magnetic field is applied in the magnetic core of a solenoid, an homogeneous 
magnetic field produces a magnetization in the core and, in this condition, the dipoles exhibit 
the same excited states with superposition between $| \ 0 \ \rangle$ and $| \ 1 \ \rangle$ even 
reversing the direction of the field from upward to downward and vice-versa. Consequently the 
states from the external dipoles of the environment follow the same state changes considering the 
existing mutual entanglement between them. The magnetic dipoles from the magnetic core and the 
environment suffer a physical (mechanical) displacement according to the magnetic field direction. 
The result of this nonlocal interaction is a net upward force where the magnetic cylinder core
placed on the ground by gravity has an upward displacement from the region with higher density 
(inside planet) to the lower density of dipoles (atmosphere and outer space). This upward force reduces 
the weight of the solenoid such as measured in our experiments. 

This effect is almost analogue to the effect that occurs in piezoelectric crystals placed on 
the ground by gravity, that suffers a shape expansion in the vertical direction and have their 
center of the mass showing an upward displacement when an electric field in the vertical 
direction is applied. However, the difference in this latter case is the existence of the 
local interaction (electric contact force between the crystal and the ground).

\subsection{Electric Dipole Forces and Clausius-Mossotti Formalism}

The same hypothesis of nonlocal interaction for magnetic dipoles seems to explain also the 
reduction of weight measured in experiments involving symmetric capacitors~\cite{Elio1,Elio2} 
even when the direction of the electric field across a dielectric is reversed by considering 
their symmetry axis with an upward direction. The electric dipoles exhibit the same excited 
states with superposition between $|$ 0 $\rangle$ and $|$ 1 $\rangle$ when the strong electric 
field is applied across a capacitor dielectric. In those experiments, the influence of the electric 
interactions with the environment around it became negligible by electric shielding implementation 
and other procedures. For this reason, the possibility of the hypothesis of the nonlocal interaction 
seems to be very likely. 

The system of two entangled electric dipoles can be well described by the Hamiltonian showed in 
Ref.~\cite{Wei}:
\begin{equation}
\hat{H} = \hbar \sum_{i=1}^{N} \omega_i \hat{S}_i^{z} + \hbar \sum_{i \ne j}^{N} \Omega_{ij} \hat{S}_i^+\hat{S}_j^-, 
\label{hamil-elec}
\end{equation}
in which each observable $\hat{S}$ represents the spin operator of a dipole in the pair and $\Omega$ is the 
electric coupling parameter between them. One can consider that there is a negligible electric coupling 
($ \Omega = 0$) between the two dipoles. The dipole moments can be oriented either along \ ( $|$ 0 $\rangle$ ) or 
against ( $|$ 1 $\rangle$ ) the external electrical fields. In this condition, there are two entangled states 
for the pair of dipoles represented by the same equations (\ref{eq4}) and (\ref{eq5}) showed 
before for the entangled magnetic dipoles.

In case of works related to the symmetric~\cite{Elio1,Elio2} and asymmetric capacitors~\cite{Elio3}, 
the equations for the force calculation have also applied a classical quantity such as the electric 
susceptibility $\chi_e$ and some analogy with the magnetic susceptibility $\chi_m$ can be taken 
in place accordingly. In other words, the electric susceptibility $\chi_e$ can be considered 
as an entanglement witness~\cite{Marcin}. Obviously all of those themes need much more detailed 
quantitative analysis.

In the 19th century, the relationship between macroscopic observables like electric susceptibility $\chi_e$ 
and individual properties of atomic or molecular electric dipoles such as polarizability $\alpha$ was firstly 
realized by Mossotti and Clausius for nonpolar materials~\cite{CM}. 

The model proposed in this work is based on the theoretical description of a set of electric dipoles, 
subjected to a high electric potential. By means of the macroscopic observables, we calculate the magnitude 
of the forces generated in capacitors by employing Clausius-Mossotti equation:
\begin{equation} 
\frac{\epsilon_r - 1}{\epsilon_r + 2} = \frac{4\pi N \alpha}{3}, 
\label{eqcm} 
\end{equation} 
in which $N$ is the number of particles per volume, $\epsilon_r$ is the relative permittivity 
and $\alpha$ is the polarizability of the atom or molecule. 
Basically, Clausius-Mossotti equation presents at the left side the macroscopic variables and at the 
right one the microscopic variables. That equation provides the density of dielectric dipoles in a dielectric 
medium, based on its relative permittivity. The utilization of the equation is worth for solids and for low 
dielectric constant. 

So, by supposing the feasibility of the quantum entanglement in the dielectric system in a microscopic scale, it is 
natural to extend the idea to enclose all the macroscopic scale in a generalized way, by considering Clausius-Mossotti 
relation applied in the calculation of the electric dipolar force $F$ induced by the quantum entanglement. 

Thus, by applying the idea to generic capacitors we conjecture that such a force can be well described by the 
empirical relation: 
\begin{equation}
F =  \frac{ 0.102 } {16 . \pi^2 }  \frac{\epsilon_r-1}{\epsilon_r+2} \frac{ A_1^2 }{ A_2 }  . \epsilon_0 . E^2 , 
\label{dip}
\end{equation}
in which the force $F$ is in units of Kgf, $A_2$ is the area of the small electrode, $A_1$ is the area of the 
large electrode, $\epsilon_r$ is the relative permittivity of the material, $\epsilon_0$ is the dielectric 
constant of the vacuum and $E$ is the electric field applied on the medium. 

The equation for the dipolar force (\ref{dip}) was postulated from its known dependence on the area of the plates, the permittivity 
and the electric field applied on the capacitor, beside the correct dimensional analysis and the idea of a direct relation between 
the microscopic and macroscopic quantities. It is an empirical formula that includes the factor of the Clausius-Mossotti relation 
$\epsilon_0 (\epsilon_r-1) / (\epsilon_r+2) $ representing the electric dipoles density in the dielectric such as implemented 
similarly to equations of dielectrophoresis forces~\cite{Pohl}. Another similarity is the implementation of the gradient of the 
squared electric field but in case of the dipolar force equation, the calculation of this gradient is implemented according 
to the relation between the areas of the electrodes also explained in Ref.~\cite{Fazi}. The numeric term 
$\left(\frac{1}{16.\pi^2}\right)$ of the dipolar force equation is adopted to determine the magnitude of 
the force and it seems to be related to the boundary conditions for the interaction between the inner 
electric dipoles and the environment. In order to obtain the corresponding empirical formula of the 
dipolar force for the symmetrical case, it is enough to do $A_1=A_2=A$, so that we obtain: 
\begin{equation}
F =  \frac{ 0.102 } {16 . \pi^2 }  \frac{\epsilon_r -1}{\epsilon_r + 2} \ \epsilon_0  A  E^2 , 
\label{cm1}
\end{equation}
as already described in Refs.~\cite{Elio1,Elio2}.

It is important to report that the term 
\begin{equation}
E_e =  A \frac{\epsilon_r -1}{\epsilon_r + 2}  \epsilon_0  E^2  
\end{equation}
in Eq.~(\ref{cm1}) amounts to the summation of energy eigenvalues of all electric dipoles in the 
dielectrics of the capacitors. 

The quantum parameters seem to be embedded in equations (9), (14) and (15) related to the calculation of the 
magnitude of nonlocal forces between dipoles and environment.
	
Theoretically, in the case of the capacitors as examples, if internal electric dipoles in the dielectric 
medium are quantum entangled among them and with the external environment, hence they can induce nonlocal 
forces that affect their own dynamics when they were subjected to the application of an intense 
electric field. A lot of different experiments from different laboratories and authors, 
including our earlier experimental works~\cite{Elio1,Elio2,Elio3} performed with symmetrical 
and asymmetrical capacitors indicate strong evidences that such an effect is possible, 
that is, the devices present anomalous variations of their weights or the existence of anomalous forces 
raised on them when they are subjected to the regime of high voltage applied. 

\section{Experimental Results}
\label{sec:2}

\subsection{Measurements on Capacitors}

In order to check our theoretical framework for both magnetic and electric dipoles, we implemented 
some experimenal measurements with two devices: capacitors and magnetic cores. We initially performed 
the measurement of the maximum weight loss of a symmetrical capacitor sample that we could obtain 
for our power supply. We consider in our experiment the capacitor sample of two parallel rounded plates 
characterized in table~\ref{tabcap}, as earlier described in Refs.~\cite{Elio1,Elio2}. 
{\small{
\begin{table}
	\centering
	\caption{Features of the first capacitor used in the experiments.}
		\begin{tabular}{|c|c|}
			\hline
			Material Plates & Aluminium\\
			\hline
			Dielectric & Plastic (Polystyrene)\\
			\hline
			Capacitance & 118.3 pF\\
			\hline
			Diameter & 8 cm\\
			\hline
			Tickness & 1.0 mm\\
			\hline
			Relative Permittivity & 2.7 \\
			\hline
			Mass & 41.154 g \\
			\hline
		\end{tabular}
	\label{tabcap}
\end{table}
}}

A maximum average weight loss of up to 220 mgf was measured when it was applied a maximum DC voltage of 20 kV. 
The measurement of the capacitor weight was made by a milligram electronic scale of 300g maximum load that 
presented very strong fluctuations. Despite the accuracy of the measurements due to the interferences, the 
results indicated in a consistent way the existence of an upward force on the capacitor when its plates 
were placed in the horizontal position. 

\begin{figure}[ht]
\begin{center}
\includegraphics[width=15.8cm]{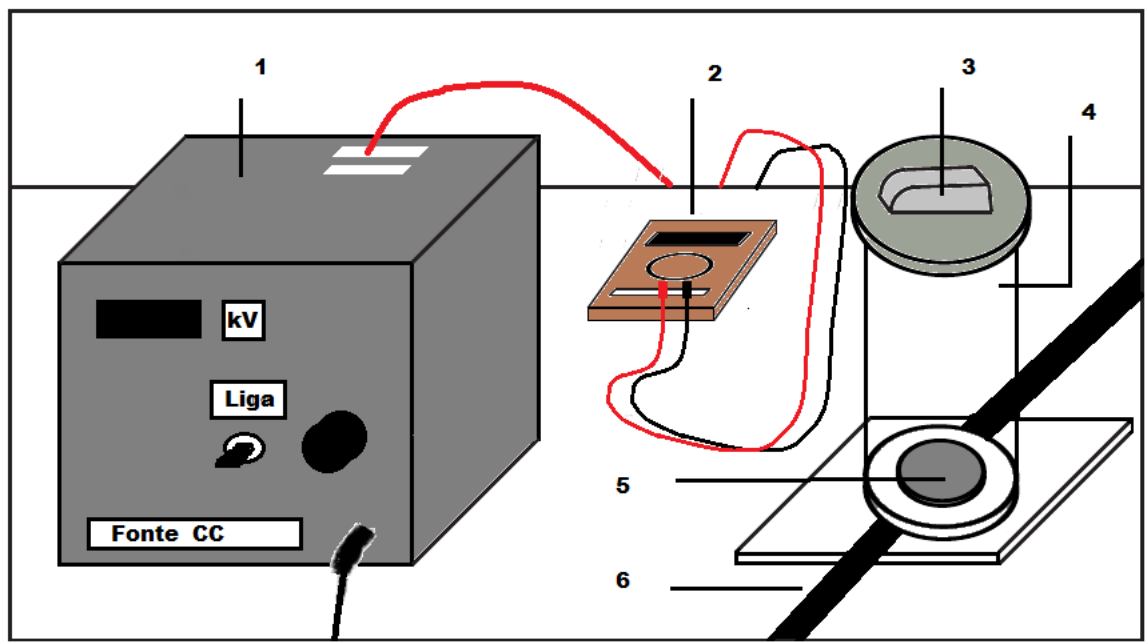}
\caption[dummy0]{Scheme of the experimental setup implemented for the measurements of weight variation of the capacitor 
devices in our experimental work. All the connections were insulated by layers of polypropylene. The z-axis of the 
accelerometer sensor was placed in the upward vertical direction and parallel to the symmetry axis of the capacitor. 
A 12 cm high cylinder of glass supported a plastic tray covered by a shielding aluminium where it was enclosed the 
accelerometer. The capacitor was connected to the high-voltage power supply and the data were collected initially 
turning on it with the power supply turned off. After a initial operation time of the accelerometer, the power 
supply was turned on during a determined operation time interval and the measurements of gravity acceleration 
were done. In the figure, we have the following devices indicated by numbers: 1-Power Supply; 2-Amperemeter; 
3-Accelerometer; 4-Glass cylinder; 5-Capacitor; 6-Insulating layers.} 
\label{fig2}
\end{center}
\end{figure}

In order to improve the accuracy of the measurements, we adopted in a second stage an experimental methodology 
already described in Refs.~\cite{Elio1,Elio2,Elio3} in which an advanced accelerometer was used as a weight 
variation sensor in order to collect a huge quantity of experimental data for the anomalous force. So we 
obtained a significant improvement in the scheme of measurements because it was possible to obtain 
automatically hundreds of measurements per second for each value of tension independently of human reading. 
In the experimental setup implemented for the measurements shown in the scheme drawn in Fig.~\ref{fig2}, we 
slowly varied the tension on our sample symmetrical capacitor and measured the average weight variation 
of the capacitor in mgf up to the maximum voltage 7~kV. 

A third experiment was different from earlier ones in relation to the geometry of the capacitor. We have 
also used in our experimental work an asymmetric capacitor device assembled with 
two parallel rounded electrodes of physical features as described in table~\ref{tabcap1}.
{\small{
\begin{table}
	\centering
	\caption{Features of the asymmetric capacitor used in the experiments.}
		\begin{tabular}{|c|c|}
			\hline
			Material Plates & Aluminium\\
			\hline
			Dielectric & Plastic (Polystyrene)\\
			\hline
			Capacitance & 3.0 pF\\
			\hline
			Small Electrode Diameter & 5.0 cm\\
			\hline
			Large Electrode Diameter & 10.0 cm\\
			\hline
			Tickness & 1.0 mm\\
			\hline
			Relative Permittivity & 2.7 \\
			\hline
			Mass & 25.0 g \\
			\hline
		\end{tabular}
	\label{tabcap1}
\end{table}
}}

Independently of possible thermal, electromagnetic, acoustic and seismic interferences one could clearly realize 
for the asymmetric case the presence of the upward force for 7 kV, by considering the standard deviations 
determined. The results were conclusive only for that highest voltage applied on the capacitor and are here summarized 
in the table \ref{table1}. The results for the asymmetrical case are also shown in the table~\ref{table1}, 
including both changes of plates positioning.

\begin{table}[h]
{\small{
\centering
\caption{Measurements of average weight variation (in mgf) of two capacitors, by considering significant time 
intervals of collected data during the operation of the power supply. The first one corresponds to the symmetrical 
capacitor, on which is applied 7~kV. In the second one, the asymmetrical case, in which was applied 6~kV in both 
position cases of the small plate, that is, parallel to the ground and positioned either up or down in the vertical 
direction, as indicated in Fig.~\ref{fig1}. In the first column we indicate the two samples used and the position 
of the large electrode. The second column indicates the operation time of the power supply; the third one shows the 
average weight variation measured by the accelerometer; In the 4th column, one reads the correspondent standard 
deviations calculated and in the last column the corresponding theoretical weight loss variations in mgf.}
\label{table1}
\begin{tabular}{|l|c|c|c|c|}
\hline
Symmetry & Operation  & Weight Variation & Error & Theoretical\\
         & Interval (s) & (mgf) & (mgf) & Value (mgf) \\
\hline 
Symmetrical & 31-60 & 36.8125 & 4.7 & 40.38\\
Symmetrical & 35-60 & 36.7344 & 3.6 & 40.38\\
Symmetrical & 40-60 & 36.7031 & 2.4 & 40.38\\  
Asymmetrical (up) & 45-55 & -251.51 & 94.3 & -185.17 \\
Asymmetrical (up) & 50-55 & -191.07 & 28.45 & -185.17 \\
Asymmetrical (down) & 45-55 & 247.78 & 62.42 & 185.17 \\ 
Asymmetrical (down) & 50-55 & 278.83 & 123.57 & 185.17 \\ 
\hline
\end{tabular}}}
\end{table}

For both symmetrical and asymmetrical cases, we implemented the referred experimental procedure of 
adopting an accelerometer configured to provice high resolution sampling, automatically obtaining 
320 samples per second of acceleration measurements for each value of voltage applied on the capacitors. 
 
The high voltage DC was provided by the power supply connected in parallel to the capacitor and 
the electric current was monitored by an amperemeter serially connected. The z coordinate of the 
accelerometer sensor was aligned according to the vertical direction and the symmetry axis of the 
capacitor. The variations of the gravity acceleration were measured by the accelerometer positioned 
on the capacitors. Such a procedure was also implemented for both positions of the asymmetric case, 
that is, with the small electrode either in upward or in downward position. The high-voltage power 
supply was turned on for some seconds after the accelerometer was switched on. In the first 
operation time interval up to 30s, the accelerometer was turned on and remained collecting data 
while the power supply was switched off; in the second step of the procedure, the power supply 
was switched on for the maximum voltage applied in each case. The difference of the average 
values of the measures between two time slots indicated an anomalous upward force in the 
symmetric capacitor and an anomalous force always pointing in the direction of the small 
electrode. It was also considered in our analysis the difference in standard deviation 
between both periods "on" and "off" and the increasing in the magnitude of the sign "on" in 
comparison with the sign "off" in terms of their amplitudes indicated a real perturbation 
caused by the anomalous force. For that high voltage applied, a relative 24~kV/mm of electric 
field was applied and the breakdown limit of the dielectric (polystyrene) was almost achieved. 
For other voltage values lower than the maximum reached in the experiment, noises from a lot 
of sources seemed to compete with the effect in a random way and the results were not conclusive. 
In fact, for such values, the level of the sign was very small and undistinguishable from the 
environment perturbations such as acoustic or thermal noises. 

It is also relevant to mention that, in the symmetrical case, we could not detect any effect 
below 3~kV, but an anomalous perturbation was observed for higher voltages from 3~kV, although 
the competition with interferences during the operation did not allow a positive conclusion on 
the presence of anomalous forces for all operation time ranges V $<$ 7~kV. In the asymmetrical 
one, we also observed perturbations for V $<$ 6~kV, but analyzing the standard deviations the 
existence of anomalous forces was not conclusive. In summary, for the voltage 7~kV applied on 
the symmetrical capacitor and 6~kV on the asymmetrical one, we really verify that the anomalous 
effect existed because the difference between two periods of average measures indicated a 
consistent upward force even changing the polarity of the capacitor and operation time 
intervals. By following that procedure, we obtain measurements that are consistent with 
the weight variation hypothesis for the higher voltage applied. Despite the weakness of 
the effect, we believe that in hypothesis it could be enhanced if we had used a device 
with high physical values, as supercapacitance~\cite{Markoulidis} or by techniques of 
optical control of capacitance through a dielectric constant~\cite{Yamasaki}. 

We realized in the results shown in the table~\ref{table1} that the symmetrical case is more stable, 
presenting smaller standard deviations. For the asymmetrical case, the variations are higher than in 
the symmetrical one, but the average value of the weight loss remained in the same order of magnitude. 
In terms of the relevance of the measurements, in all of the cases, for hundreds of data per second 
there was the detection of the weight variation and with same order of magnitude for all operation 
time ranges considered in the experiment, corroborating the existence of the anomalous force. 

\subsection{Measurements on Magnetic Cores}

Beside our experimental measurements concerning to apparent weight losses in capacitors, we also 
implemented an experimental scheme in order to measure anomalous forces in magnetic cores. 
\begin{table}[h]
{\small{
\centering
\caption{Features of the devices used in the experimental scheme for measuring anomalous forces in 
magnetic cores. }
\label{table2}
\begin{tabular}{|l|l|}
\hline
Device & Features \\
\hline\hline
Digital Ampere Meter & UNI-T model UT30B; Current Range 200$\mu$A $\sim$ 10A; \\
 & Resolution 100nA (for 200 $\mu$A DC current); \\ 
 & 3 $\frac{1}{2}$ digits LCD Display \\
\hline
Digital Scale & BEL model S303; Capacity 310g; Readability 0.001g; \\ 
 & Linearity $\pm$0.003g; Reproducibility 0.0006g; \\ 
 & Power Supply 115/230 VAC $\pm$ 15\%; LCD display \\
\hline
1st Battery & +3 VDC; Panasonic; CR1620 model; 75mAh \\
\hline
2nd Battery & +6 VDC; Eveready; Super Heavy Duty 1209; 11Ah \\
\hline
Variable Resistor Rheostat & 120 Ohm; 10\%; 200 W \\
\hline
1st Magnetic Coil & 1 cm diameter; 2.7 cm length, 192 loops; \\ 
 & Iron Alloy cylinder core \\
\hline
2nd Magnetic Coil & 1.6 cm diameter; 5 cm length; 252 loops; \\
 & Ferrite cylinder core \\
\hline
Neodymium Magnet & 1.4 cm diameter, 1.4 cm length; 1.17 Tesla; \\
 & NdFeB cylinder core \\
\hline
\end{tabular}}}
\end{table}  
 
\begin{figure}[h]
\begin{center}
\includegraphics[width=8cm]{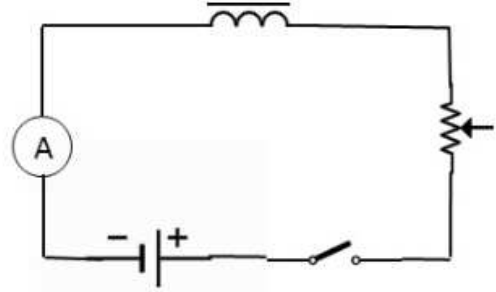}
\caption[dummy0]{Scheme of the experimental circuit mounted for the measurements of the weight losses (in mgf) 
of the magnetic core.} 
\label{fig3}
\end{center}
\end{figure}

\begin{figure}[h]
\begin{center}
\includegraphics[width=15cm]{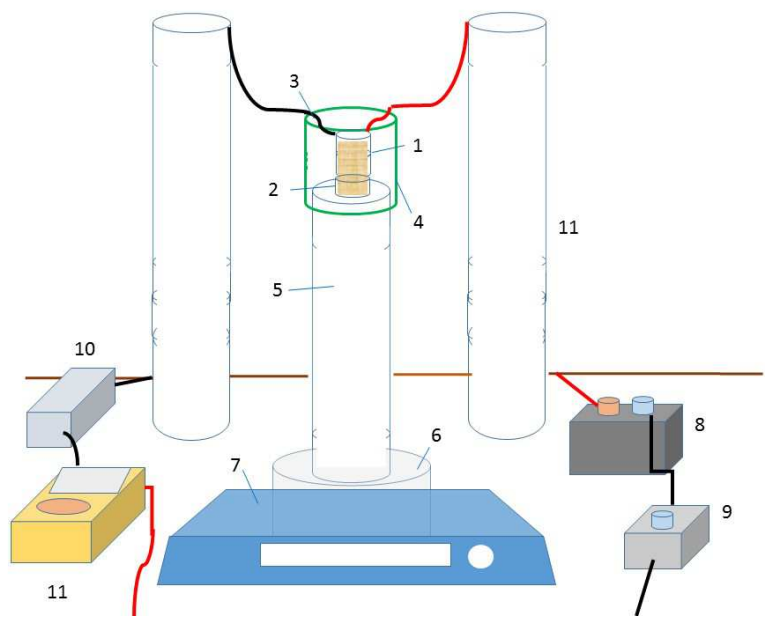}
\caption[dummy0]{Schematic diagram of the experimental setup used for the measurements of the weight losses (in mgf) of the 
magnetic core. In the figure, we have the following devices indicated by numbers: 1-Magnetic coil; 
2-Foan of plastic; 3-Wire; 4-Hollow cylinder; 5-Cylinder of cardstock; 6-Circular platform; 7-Digital scale; 8-Battery; 
9-On/Off switch; 10-Variable resistor rheostat; 11-Digital ampere meter; 12-Cylinder of cardstock.} 
\label{fig4}
\end{center}
\end{figure}

In table \ref{table2}, we show the main features and technical information of the devices used in our experimental assembly. The data shown also concerns to the instruments that we have used in the measurements by using the test magnetic coil devices. In order to measure the anomalous force in the solenoid, we have used the magnetic coils and instruments described in table \ref{table2} according to the electric circuit mounted in Fig.~\ref{fig3} and the schematic diagram of the experimental assembly shown in Fig.~\ref{fig4}. 

In our experimental measurements, we used a magnetic core containing three different magnets: two magnetic coils and one neodymium magnet. The first magnetic coil had dimensions 1 cm diameter, 2.7 cm length and 192 loops around a iron allow cylinder core. The second one had 1.4 cm diameter, 5 cm length, 252 loops and a ferrite cylinder core. The setting was also constituted by a neodymium magnet of 1.4 cm diameter, 1.4 cm length, 1.17 T and a NdFeB cylinder core. 

In the experimental set, we used two batteries of 3 DVC and 6 DVC and varied the value of a rheostat of 120 Ohm and 200 W. The measurements of anomalous forces on the magnetic apparatus indicated the existence of a real perturbation on that system. Systematic 
sequences of measurements by using the a digital scale BEL model S303 with capacity 310~g were implemented by applying a high voltage 
by means of a power supply of 115/230 VAC $\pm$ 15\%. The current was controlled by a ampere meter UNI-T model UT30B with current range 
200~$\mu$A $\sim$ 10~A and resolution 100~nA (for 200~$\mu$A DC current). Our measurements do indicate the raising of a anomalous 
force on the setting, whose results are presented in table~\ref{tab5}.  

We performed some experimental measurements with the objective of verifying the maximum weight loss of the magnetic coil samples. Firstly the magnetic force of the 1st magnetic coil with 1 cm diameter, 2.7 cm length, 192 loops and iron alloy cylinder core in the vertical position was measured with its symmetry axis aligned according to the symmetry axis of the circular stainless steel platform of the electronic scale when a 2.1 A current flow was applied through the coil. The distance between the platform of the digital scale and the cylinder core was 2 cm. The readability of the digital scale was 0.001 g and its capacity was 310 g. Under such conditions, a 2 gf force was measured. 

In the measurements involving the magnetic devices, we did not have the value of the magnetic susceptibility of the core, so that we needed to implement an experimental procedure for the determination of the magnetic field. The magnetic field B could be calculated using the known 
Maxwell's pulling force formula~\cite{Windt}:
\begin{equation}
F=\frac{SB^2}{2\mu_0}, 
\label{force}
\end{equation}
where $F$ is the magnetic force, $S$ is the surface area of the cylinder core, $B$ is the magnetic field and $\mu_0$ is the vacuum permeability. A 1.072 T of magnitude of the magnetic field was calculated and its value is below the saturation range of the iron 
alloy, that is between 1.6 T and 2.2 T. Some measures were performed and the magnetic interaction between the coil and the platform 
of the digital scale became negligible for distances higher than 8 cm.

Afterwards the same magnetic coil was positioned on the top of a 31.5 cm high cylinder of cardstock with their symmetry axis aligned mutually in the vertical direction. The cylinder of cardstock was positioned over the center of the circular platform of the digital 
scale and its length ensured an adequate distance between the platform of the digital scale and the magnetic coil in order to 
strongly reduce the magnetic interaction. Two wires provided an electric connection of the magnetic coil and two other cylinders of cardstock supported them such as showed in the Fig.~\ref{fig4}. A hollow cylinder of thin iron housed the magnetic coil as a magnetic shield in order to reduce 10 times its outer field. The efficiency of this magnetic shield was proved according to the reduction of 
the magnetic force of a neodymium magnet class N35, 1.4 cm diameter, 1.4 cm length and 1.17 T magnetic field. A foan of plastic was installed below the magnetic coil to reduce any possible effect of the shape magnetic core variation such as a magnetic constriction. 
The magnetic coil was connected in series with the on/off switch, battery of 6 V, variable resistor rheostat and 
ampere meter, as showed in the schematic diagram of the Fig.~\ref{fig4}. 

A 2.1 A electric current magnitude (average) was measured by the ampere meter and controlled by the rheostat when the switch was 
turned on. In this condition, the weight loss of the setup varied between 15 and 20 mgf including magnetic coil, wires, hollow 
cylinder, foan of plastic and cylinder cardstock according to the average of maximum values of several measurements performed by 
the digital scale. The initial weight value of the setup (149.097 gf) was recovered when the switch was turned off again. The 
direction of the electric current was changed but the weight loss of the setup remained the same. It is also important to note 
that weight loss value remained the same despite the fact that some devices such as magnetic shield (hollow cylinder of thin 
iron) and foan of plastic were later removed from the setup when new measurements were performed. There were no changes 
in the weight loss readings even replacing the standard type of wires from cooper to other metallic materials for electric connections of the magnetic coil. In other words, the metallic constriction of the wires regarding the temperature and the electric current flow seemed extremely low in terms of possibility to affect the weight measurements also especially considering the catenary of the suspending wires.

The experimental features seem to exclude any explanation based on magnetic field interaction between the magnetic coil and the environment (including the digital scale) regarding the weight loss and they reinforce the explanation showed in the theoretical description section related to the nonlocal force.

Considering a 2.1 A electric current and a 1.072 T magnetic field calculated before it is possible to calculate the magnitude of the nonlocal force for the 1st magnetic coil regarding the Eq.~(\ref{forcedip1}). The value of the force magnitude obtained is 22.84 mgf. This theoretical value is close to the experimental value of the range between 15 mgf and 20 mgf.

In order to improve our analysis, we performed other experimental measurement using, as showed in table~\ref{table2}, a 2nd magnetic coil  with 1.6 cm diameter, 5 cm length, 252 loops and ferrite cylinder core. It was performed the measurements in the same setup mentioned before just replacing the magnetic coil 1 for the magnetic coil 2. The magnitude of the magnetic field of 0.3 T was calculated for an electric current of 1.6 A via a measurement of the magnetic force between the magnetic coil and the platform of the digital scale such as the procedure already mentioned before. This 0.3T magnetic field magnitude value is close to the ferrite magnetic saturation range between 
0.2 T and 0.5 T. In other words, the magnetic core was probably saturated or close to this condition.

The average of maximum values of the several measurements of the weight loss of the same setup using the 2nd magnetic coil was the range between 8 and 10 mgf regarding the electric current of 1.6 A and the magnetic field 0.3T. Considering the same parameters, the nonlocal force magnitude calculated via Eq.~(\ref{forcedip1}) is 7.79 mgf and this value is really close to the experimental values of the range mentioned before.

Considering the same setup of the magnetic coil 2 except that the battery was replaced to one of 3 V, it was performed new measurements and a weight loss between 1 and 2 milligrams was detected regarding the 0.17 A electric current flowing in the circuit when the on/off switch was turned on accordingly. A 0.828 mgf of theoretical value of the weight loss is also close to the experimental values of the range regarding the 0.3 T magnitude of the magnetic field for a no saturated ferrite magnetic core.

A summary of the experimental results is shown in table~\ref{tab5}.
\begin{table}
{\small{
	\centering
	\caption{Comparison between several measurements of the weight loss considering different magnetic coils and parameters as average electric current and magnetic field magnitude. There is a 3mg linearity error of the digital scale and therefore the theoretical values are in accordance with the range of measurement values. }
		\begin{tabular}{|l|l|l|l|l|}
\hline
Experimental	& Magnetic Field  &	Average Electric	& Weight Loss & Theory  \\
Setup                    & Magnitude       & Current           & Range        & Forecast  \\  
  & (T) & (A) & (mgf) & (mgf) \\
\hline
Magnetic coil 1 	& 1.072 	            & 2.1 	  & 15 $\sim$ 20 	 & 22.84 \\
(iron alloy core) & & & & \\
\hline
Magnetic coil 2 	  & 0.3  	& 1.6 	  & 8 $\sim$ 10 &	7.79 \\
(ferrite core) & (saturation) & & & \\
\hline
Magnetic coil 2 	  & 0.3 	              & 0.17 	& 1 $\sim$ 2 	 &0.828 \\
(ferrite core) & & & & \\
\hline
		\end{tabular}
	\label{tab5}
}}
\end{table}

The procedures adopted in the experimental setups seem to exclude the outer magnetic interaction and other effects such as magnetic constriction of the core or variation of the shape of the wire conductors regarding the thermal effect of the electric current flow as the cause of the weight loss of the setups.

The values of the experimental results are close to the theoretical values where a coupling between the magnetic dipoles and the environment via quantum entanglements is taken in consideration.

The authors plan to improve the future experiments and obtain a curve of the weight variation versus the electric current applied in order to better analyze the saturation effect and the standard deviation.

From the exposed results, we conclude that our theoretical proposal is consistent in explaining our experimental 
results and or the anomalous effects in magnetic and electric devices, based on the macroscopic observables as 
manifestations of the microscopic generalized quantum entanglement among dipoles and environment. 
\section{Conclusion} 
In this work, we present our experimental and theoretical investigations concerning to the 
existence of an anomalous force on symmetrical and asymmetrical capacitors, operating in high 
voltage, and on magnetic cores operating in high magnetic fields. Our investigation was 
motivated by a lot of earlier experimental works described in the literature performed by 
other different authors. The existence of anomalous forces on magnetic cores was also 
investigated in our hitherto unpublished experiments. The detected anomalous forces 
seem to be not explained by electric or magnetic interactions between devices and 
environment considering the procedures adopted accordingly. This fact reinforces our 
hypothesis related to an nonlocal interaction, considering a preexisting state of 
generalized quantum entanglement between all particles. Considering some macroscopic 
observables such as the magnetic and electric susceptibility involving quantum 
entanglements between magnetic or electric dipoles, we explained how the magnitude of 
the so called anomalous forces can be calculated via equations using classical quantities. 
In fact, the theoretical results show that such a concept can explain with good accuracy 
the majority of the experimental data. We intend in a next step to study the application 
of the preexisting condition of quantum entanglement to investigate its relation to the 
gravity and inertia. We also aim to investigate how new materials or other possible 
configurations with the devices analyzed or new setup possibilities could enhance the 
interaction so that one can use it in technological applications as electric propulsion. 
Besides, it would be convenient also to report that we are elaborating other works based 
on a new experimental setup involving the detection of higher magnitudes of the anomalous 
forces and the effect of induction of nonlocal forces at distance, having high voltage 
capacitors as other source devices.


\section*{References}

\begin{thebibliography}{99}

\bibitem{Brown1} Brown T T 1928 A Method of and an Apparatus or Machine for Producing Force or Motion 
U.K. Patent No. 00.311 

\bibitem{Brown2} Brown T T 1965 Electrokinetic Apparatus U.S. Patent 3.187.206 

\bibitem{Williams} Williams P E 1983 The Possible Unifying Effect of the Dynamic Theory Rept LA-9623-MS Los Alamos Scientific Lab. Los Alamos 
NM 

\bibitem{Manitoba} Buehler D R 2004 Exploratory Research on the Phenomenon of the Movement of High Voltage Capacitors 
{\it Journal of Space Mixing} Vol.~2 1

\bibitem{Mahood} Mahood T L 1999 Propellantless Propulsion: Recent Experimental Results Exploiting Transient Mass Modification, 
CP458 Space Technology and Applications International Forum, edited by Mohamed S. el-Genk 

\bibitem{Musha0} Musha T 2014 Theoretical Explanation of the Biefeld-Brown Effect, URL: http://www.ovaltech.ca/electrogravity.html

\bibitem{Musha} Musha T 2008 Explanation of dynamical Biefeld-Brown Effect from the standpoint of ZPF field 
{\it JBIS} Vol.~61 379-384

\bibitem{Fazi} Bahder T B and Fazi, C 2005 Force on an Asymmetric Capacitor {\it Electrogravitics II} (Integrity Research Institute, 
Washington, 28-59) 

\bibitem{Woodward1} Woodward J F 1991 Measurements of a Machian Transient Mass Fluctuation 
{\it Found. Phys. Lett.} Vol.~4 407-423  

\bibitem{Woodward2} Woodward J F 1992 A Stationary Apparent Weight Shift From a Transient Machian Mass Fluctuation 
{\it Found. Phys. Lett.} Vol.~5 425-442 

\bibitem{Woodward3} Woodward J F 1996 A Laboratory Test of Mach's Principle and Strong-Field Relativistic Gravity 
{\it Found. Phys. Lett.} Vol.~9 247-293 

\bibitem{Modanese} Modanese G 2013 A comparison between the YBCO discharge experiments by E. Podkletnov and C. Poher, and their theoretical interpretations {\it Appl. Phys. Res.} Vol.~5 No.~6 59-73  

\bibitem{Martins} Martins A A and Pinheiro 2011 M J On the propulsive force developed by asymmetric capacitors in a vacuum 
{\it Phys. Proc.} 20 112-119  

\bibitem{Canning1} Canning F X, Melcher C and Winet E 2004 Asymmetrical Capacitors for Propulsion (NASA/CR-2004-213312, 1-16)

\bibitem{Canning2} Canning F X, Cole J, Campbell J and Winet E 2004 The ISR Asymmetrical Capacitor Thruster, Experimental Results and Improved Designs (40th AIAA/ASME/SAE/ASEE Joint Propulsion Conference and Exhibit Fort Lauderdale, Florida, 11-14) 

\bibitem{NASA} NASA 2001 Apparatus and Method for Generating a Thrust Using a Two Dimensional Asymmetrical Capacitor Module U S Patent 
6,317,310; Campbell J W 2002 Apparatus for Generating Thrust Using a Two Dimensional Asymmetrical Capacitor Module U S Patent 
6,411,493 

\bibitem{Poher} Poher C and Poher D 2011 Physical Phenomena Observed during Strong Electric Discharges into Layered Y123 Superconducting Devices at 77 K {\it Applied Physics Research} Vol.~3 No.~2  

\bibitem{FungChung} Chung C F and Li W J 2007 Experimental Studies and Parametric Modeling of Ionic Flyers (IEEE/ASME international 
Conference on Advanced Intelligent Mechatronics, 1-6) 
 
\bibitem{Elio1} Porcelli E B and Filho V S 2016 On the Anomalous Forces in High-Voltage Symmetrical Capacitors {\it Phys. Essays} 
Vol. 29 2-9   

\bibitem{Elio2} Porcelli E B and Filho V S 2015 On the Anomalous Weight Losses in High-Voltage Symmetrical Capacitors 
arXiv:1502.06915 

\bibitem{Elio3} Porcelli E B and Filho V S 2016 Characterisation of anomalous asymmetric high-voltage capacitors {\it IET Science, 
Measurement \& Technology} Vol. 10 Issue 4 383–388 
 
\bibitem{Ianconescu} Ianconescu R, Sohar D and Mudrik M 2011 An analysis of the Brown–Biefeld effect J. Electrostatics Vol. 69 Issue 6 
512–521  

\bibitem{Puthoff98} Puthoff H E 1998 Can the vacuum be engineered for spaceflight applications? Overview of theory and experiments  
{\it J. Sci. Exploration} Vol. 12 295  

\bibitem{Puthoff02} Puthoff H E, Little S R and Ibison M 2002 Engineering the Zero-Point Field and Polarizable Vacuum for Interstellar Flight 
JBIS Vol. 55 137-144  

\bibitem{Li} Li L, Lee S J, Kim W and Kim D 2015 An empirical model for ionic wind generation by a needle-to-cylinder dc corona discharge 
{\it J. Electrostatics} Vol. 73 125-130 

\bibitem{Zhao} Zhao L and Adamiak K 2006 EHD gas flow in electrostatic levitation unit J. Electrostatics 64 639-645  

\bibitem{Zhang} Zhang Y, Liu L, Chen Y and Ouyang J 2015 Characteristics of ionic wind in needle-to-ring corona discharge 
{\it J. Electrostatics} Vol. 74 15-20  
 
\bibitem{Puthoff89} Puthoff H E 1989 Gravity as a zero-point-fluctuation force 
{\it Phys. Rev. A} 39 2333 

\bibitem{Puthoff91} Puthoff H E 1989 On the Source of Vacuum Electromagnetic Zero-Point Energy 
{\it Phys. Rev. A} 40 4857; Errata and Comments 1991 {\it Phys. Rev. A} 44 3382-3385  

\bibitem{Puthoff90} Puthoff H E 1990 Everything for Nothing {\it New Sci.} 127 52   

\bibitem{Haisch} Haisch B, Rueda A and Puthoff H E 1994 Inertia as a zero-point field Lorentz force  
{\it Phys. Rev. A} 49 678; Rueda A and Haisch B 1998 Inertia as reaction of the vacuum to accelerated motion 
{\it Phys. Lett. A} 240 115 

\bibitem{Antano} Antano M A 2012 Biefeld-Brown Effect and Space Curvature of Electromagnetic Field arXiv:1004.0810v2  

\bibitem{Moller} Christensen E A and M\o ller P S 1967 Ion-Neutral Propulsion in Atmospheric Media 
{\it AIAA J.} Vol.~5 No. 10 1768–1773 

\bibitem{Tajmar} Tajmar M 2004 Biefeld-Brown Effect: Misinterpretation of Corona Wind Phenomena
{\it AIAA J.} Vol.~42 Nr. 2 315-318 

\bibitem{Clausius} Clausius R 1879 Die Mechanische W\"armlehre Vol. II (Vieweg-Verlag, Brunswick, Germany)

\bibitem{Mossotti} Mossotti O F 1850 Mem. di Mathem. e Fisica in Modena 24 II 49 

\bibitem{Rysselberghe} Rysselberghe P V 1932 Remarks concerning the Clausius~-~Mossotti Law 
{\it J. Phys. Chem.} Vol. 36 No. 4 1152-1155    

\bibitem{EPR} Einstein A, Podolski B and Rosen N 1935 Can Quantum-Mechanical Description of Physical Reality Be Considered Complete? 
{\it Phys Rev.} Vol. 47 777-780

\bibitem{Bohm} Bohm D 1951 Quantum Theory (Dover Publication)  

\bibitem{Bell} Bell J S 1987 Speakable and Unspeakable in Quantum Mechanics (Cambridge, University Press)

\bibitem{Zeilinger} Zeilinger 1999 A Experiment and the foundations of quantum physics {\it Rev. Mod. Phys.} 71 S288  

\bibitem{Lengeler} Lengeler B 2007 Coherence Appl. Phys. A 87 585-592  

\bibitem{Penrose} Penrose R 2004 Road to Reality (Jonathan Cape, 1st edition, London) 

\bibitem{Buniy} Buniy R V and Hsu S D H Everything is Entangled 2012  
{\it Phys. Lett. B} Vol. 718 233-236  

\bibitem{Elio4} Porcelli E B A Theoretical Insight into the Connection between the Gravity and the 
Generalized Quantum Entanglements Open Science Repository Physics e45011836  

\bibitem{Amico} Amico L, Fazio R Osterlosh A and Vedral V 2008 Entanglement in Many-Body Systems  
{\it Rev. Mod. Phys.} Vol. 80 517 

\bibitem{Vedral1} Vedral V 2010 Decoding Reality: the Universe as Quantum Information (Oxford University Press) 

\bibitem{Vedral} Vedral V 2011 A Vida em um Mundo Qu\^antico {\it Sci. Am. Br.} Vol. 110 30 

\bibitem{Ghosh} Ghosh S {\it et al.} 2003 Entangled Quantum State of Magnetic Dipoles {\it Nature} Vol. 425 48  

\bibitem{Marcin} Wie\'sniak M, Vedral V and Brukner C 2005 Magnetic Susceptibility as a Macroscopic Entanglement Witness  
{\it New J. Phys.} Vol. 7 258 

\bibitem{Wu} Wu L-A, Bandyopadhyay S, Sarandy M S and Lidar D A 2005 Magnetic Susceptibility as a Macroscopic Entanglement Witness 
{\it Phys. Rev. A} Vol. 72 032309 

\bibitem{Toth} Toth G 2005 Entanglement Witnesses in Spin Models {\it Phys. Rev. A} Vol. 71 010301(R)  

\bibitem{Dowling} Dowling M R, Doherty A C and Bartlett S D 2004 Energy as an Entanglement Witness for Quantum Many-Body Systems  
{\it Phys. Rev. A} Vol. 70 062113  

\bibitem{Wang} Wang X 2002 Thermal and Ground-State Entanglement in Heisenberg XX Qubit Rings {\it Phys. Rev. A} Vol. 66 034302  

\bibitem{Hao} Hao X and Zhu S 2007 Entanglement in a Quantum Mixed-Spin Chain {\it Phys. Lett. A} Vol. 366 206-210  

\bibitem{Chakraborty} Chakraborty T {\it et al.} 2012 Quantification of Entanglement from Magnetic Susceptibility for a Heisenberg Spin 1/2 System {\it Phys. Lett. A} Vol. 376 2967-2971  

\bibitem{Craig} Craig D K 2016 Electro-Magneto-Mechanics, Rensselaer Polytechnic Institute, URL:~  
http://www.multimechatronics.com/images/uploads/mech$\_$n/~\\~Electromechanics.pdf  

\bibitem{Wei} Wei Q, Kais S and Chen Y P 2010 Entanglement Switch for Dipole Arrays {\it J. Chem. Phys.} Vol. 132 121104  

\bibitem{CM} Markov K Z 1999 Elementary Micromechanics of Heterogeneous Media Chapter 1 in: Heterogeneous Media: Modelling and 
Simulation (Edited by K. Z. Markov and L. Preziosi, Birkhauser Boston, 1-162) 

\bibitem{Pohl} Pohl H A 1978 Dielectrophoresis the Behavior of Neutral Matter in Nonuniform Electric Fields (Cambridge University Press, Cambridge)

\bibitem{Markoulidis} Markoulidis F, Lei C and Lekakou C 2013 Fabrication of high-performance supercapacitors based on transversely oriented  carbon nanotubes {\it Appl. Phys. A} 111 227-236 

\bibitem{Yamasaki} Yamasaki K, Juodkazis S, Lippert T, Watanabe M, Matsuo S and Misawa H 2003 Dielectric breakdown of rubber materials by femtosecond irradiation {\it Appl. Phys. A} Vol. 76 325-329  

\bibitem{Windt} Windt C W, Soltner H, Dusschoten D V and Bl\"umler P 2011 A portable halbach magnet that can be opened without force: the NMR-CUFF {\it J. Magnetic Resonance} Vol. 208 27-33  

\end{thebibliography}
\end{document}